\documentclass[reprint,aps,showpacs,preprintnumbers,amsmath,amssymb,]{revtex4-1}

\usepackage[dvips]{graphicx}
\usepackage{dcolumn}
\usepackage{bm}
\usepackage{array}

\begin{document}

\title{Violation of detailed balance accelerates relaxation}

\author{Akihisa Ichiki}
\email{ichiki@gvm.nagoya-u.ac.jp}
\affiliation{Green Mobility Collaborative Research Center, Nagoya University, Furo-cho, Chikusa-ku, Nagoya 464-8603, Japan}
\author{Masayuki Ohzeki}
\email{mohzeki@i.kyoto-u.ac.jp}
\affiliation{Department of Systems Science, Graduate School of Information, Kyoto University, Yoshida-Honmachi, Sakyo-ku, Kyoto 606-8501, Japan}

\date{\today}

\begin{abstract}
Recent studies have experienced the acceleration of convergence in Markov chain Monte Carlo methods implemented by the systems without detailed balance condition (DBC).  However, such advantage of the violation of DBC has not been confirmed in general.  We investigate the effect of the absence of DBC on the convergence toward equilibrium.  Surprisingly, it is shown that the DBC violation always makes the relaxation faster.  Our result implies the existence of a kind of thermodynamic inequality that connects the nonequilibrium process relaxing toward steady state with the relaxation process which has the same probability distribution as its equilibrium state.  
\end{abstract}

\maketitle

Markov chain Monte Carlo (MCMC) methods have been widely employed to obtain the ensembles of physical quantities in systems with large degrees of freedom, such as spin glasses~\cite{Ogielski}, protein folding problems~\cite{Mitsutake}, and glass transitions~\cite{Yamamoto}.  Since Metropolis {\it et al}. successfully employed MCMC to liquid system~\cite{Metropolis}, various extensions have been invented.  Such variants all focus on their fast convergence toward the target distribution, which is often referred to as the equilibrium distribution.  One way to improve the convergence is the use of extended ensemble methods such as the Wang-Landau method~\cite{WangLandau} and the replica exchange Monte Carlo method~\cite{HukushimaNemoto}.  The alternative is the clustering methods such as the Swendsen-Wang algorithm~\cite{Swendsen} and the Wolff algorithm~\cite{Wolff}.  Recent progress in MCMC is achieved in different ways by several independent studies~\cite{SuwaTodo, Turitsyn, Fernandes}.  They employed the transition probability without detailed balance condition (DBC).  Following the success of DBC-violating algorithms, MCMC without DBC has become of interest for investigations of fast converging sampling methods~\cite{SakaiHukushima}.  

Although DBC is the sufficient condition where the system probability converges toward the target distribution after long-time relaxation from arbitrary initial distribution, the systems without DBC are also allowed to converge toward the target distribution when the balance condition (BC) is satisfied.  Exploiting BC, Suwa and Todo aimed to reduce the reject rate, which is expected to dull the convergence, by violating DBC to sweep the system states faster~\cite{SuwaTodo}.  Besides such intuitive explanation of the advantage of DBC violation, a rigorous proof has not been obtained.  A rigorous relation between reject rate and convergence is obtained only in the systems with DBC by Peskun~\cite{Peskun}.  However numerical studies on MCMC without DBC~\cite{Turitsyn, Fernandes, SakaiHukushima} subsequent to Suwa and Todo also show the fast convergence.  Thus it is expected that the violation of DBC always accelerates the convergence toward the target distribution.  In this paper, we give a rigorous proof of the advantage of violation of DBC.  

Although DBC is expected to hold in equilibrium states, the violation of DBC plays a crucial role in nonequilibrium thermodynamics, especially the current studies on steady state thermodynamics~\cite{HatanoSasa, HaradaSasa}.  The violation of the fluctuation-response relation in nonequilibrium systems has discussed in Ref.~\cite{HaradaSasa}.  Since the fluctuation-response and fluctuation-dissipation relations underlie the relaxation processes, the degree of the violation of DBC is considered to affect the macroscopic system dynamics.  The acceleration of convergence of the system without DBC in MCMC exploits nonequilibrium properties.  In this paper, we find that the nonequilibrium relaxation speed in the system without DBC is bounded by that in the system with DBC, which relaxes towards equilibrium.  The result suggests the existence of a kind of thermodynamic inequality.   

The violation of DBC can be introduced as the asymmetry of the transition rate as detailed below.  From a different perspective of stochastic quantization~\cite{Trotter, Suzuki, Nelson, Parisi}, classical systems as implemented in MCMC without DBC are equivalent to quantum systems governed by non-Hermitian Hamiltonians.  Recent studies on the effect of non-Hermitian Hamiltonians in quantum systems~\cite{Hatano, Efetov, Goldsheid, Nelson1998, Mudry} have found that delocalization occurs owing to their sensitivity to the boundary conditions and non--zero current~\cite{Hatano2}.  The delocalization affects the ergodicity of the system, which is closely related to the convergence speed in MCMC.  In addition, the non-Hermitian Hamiltonian is also exploited in several fast solvers of optimization problems by use of quantum nature as reported in Refs.~\cite{Ohzeki, Ohzeki2, Nesterov}.  In this paper, we aim at understanding the effect due to such asymmetric nature arising in various classical and quantum systems via studies on the convergence of MCMC without avoiding essential features of the asymmetric nature.   

We consider the irreducible Markov process described by the following master equation: 
$dP_i(t)/dt=\sum_{j=1}^{N}q_{ij}P_j(t)-\sum_{j=1}^N q_{ji}P_i(t)$, 
where $P_i(t)$ is the probability of state $i$ at time $t$, $q_{ji}$ is the transition rate from state $i$ to state $j$, and the sum is taken over all $N$ states.  In order to ensure the relaxation toward probability $\pi_i$, i.e., $P_i(t)\to\pi_i$ after long-time relaxation, we impose BC as 
$0=\sum_{j=1}^N q_{ij}\pi_j-\sum_{j=1}^N q_{ji}\pi_i$.  
Using BC, the master equation is rewritten as 
\begin{eqnarray}
\dfrac{d R_i(t)}{dt}=\displaystyle\sum_{j=1}^N W_{ij}R_j(t),\label{eq:masterBC}
\end{eqnarray}
where $W_{ij}=\pi^{-1/2}_i q_{ij}\pi^{1/2}_j$ for $i\neq j$, $W_{ii}=-\sum_{j(\neq i)}q_{ji}$, and $R_{i}(t)=P_i(t)/\sqrt{\pi_i}$.  The largest eigenvalue of $W$ is guaranteed to be zero by the Perron-Frobenius theorem.  Hereinafter we assume that $W$ has $N$ eigenvectors so that a linear combination of them represents a probability distribution.  The eigenvectors corresponding to nonzero eigenvalues generate the deviation from the target distribution.  Because of their negative real parts of eigenvalues, $R_i(t)$ relaxes toward $\sqrt{\pi_i}$, i.e., $P_i(t)\to\pi_i$ for all $i$ for arbitrary initial condition.  Note that $W$ plays the role of Hamiltonian on the foundation of stochastic quantization, in which the quantum system corresponding to the equivalent classical system is considered and vice versa~\cite{Trotter, Suzuki, Nelson, Parisi}.  The conservation of probability and BC are represented in terms of $W$ respectively as 
\begin{eqnarray}
\displaystyle\sum_i\sqrt{\pi_i}W_{ij}=0,\qquad \sum_j W_{ij}\sqrt{\pi_j}=0.\label{eq:BCandProbConserv}
\end{eqnarray}
Equation (\ref{eq:BCandProbConserv}) ensures that the Markov process characterized by $S=(W+W^{\rm T})/2$, in which DBC is satisfied, converges toward the equilibrium $\pi_i$ same as $W$, i.e., BC $\sum_j S_{ij}\sqrt{\pi_j}=0$ holds, where $W^{\rm T}$ denotes the transpose of $W$.  

Let us consider the relaxation of the ensemble of arbitrary physical quantity $\langle f\rangle=\sum_i f_i \sqrt{\pi_i} R_i$, which is described by the following equation:  
$d\langle f\rangle/dt=\sum_{i,j}f_i \sqrt{\pi_i}W_{ij}R_j(t)$, 
where $f_i$ is the realization of $f$ depending on the microstate $i$.  According to this equation, the relaxation time of arbitrary physical quantity is governed by the eigenvalues of $W$.  In particular, the relaxation time is dominated by the second-largest real part of the eigenvalue.  If and only if DBC $q_{ij}\pi_j=q_{ji}\pi_i$ holds, $W$ is symmetric and its eigenvalues, except zero, are all real negative.  In order to understand the effect of the violation of DBC in $W$ systematically, we decompose $W$ into the sum of its symmetric part $S$ and anti--symmetric part $\Gamma$ as $W=S+\Gamma$.  

The main claim of this paper is that ${\rm Re}\,\lambda_2^{W}-\lambda_2^S\le 0$ always holds with a fixed $S$, where $\lambda_n^W$ and $\lambda_n^S$, respectively, represent the eigenvalues of $W$ and $S$ ordered by their real parts as ${\rm Re }\,\lambda_1^W\ge{\rm Re }\,\lambda_2^W\ge\cdots\ge{\rm Re }\,\lambda_N^W$ and $\lambda_1^S\ge\lambda_2^S\ge\cdots\ge\lambda_N^S$.  Here $\lambda_1^W=\lambda_1^S=0$.  The inequality ${\rm Re }\,\lambda_2^{W}\le\lambda_2^S$ ensures that the relaxation is quickened by the violation of DBC.  The proof consists of two steps: We show that (I) $d\det(\lambda E_N-W)/d\lambda|_{\lambda=0}\ge d\det(\lambda E_N-S)/d\lambda|_{\lambda=0}$, where $E_N$ is an identity matrix of order $N$.  Next, we show that, (II) $|\det(\lambda E_N-W)/\lambda| \ge\det({\rm Re }\lambda E_N-S)/\lambda$ for complex $\lambda$ satisfying $\lambda_2^S<{\rm Re }\lambda<\lambda_1^S=0$.  If $|\det\Gamma|$ is small, $\lambda_2^W$ is expected to be real and perturbatively shifted from $\lambda_2^S$.  The combination of propositions (I) and (II) states that the shift is always in the negative direction, which ensures the faster convergence.  As $|\det\Gamma|$ increases, $\lambda_2^W$ is expected to decrease and finally become complex.  Proposition (II) confirms that such complex eigenvalues cannot deteriorate the convergence, i.e., ${\rm Re }\,\lambda_2^W$ cannot shift right to $\lambda_2^S$, even when $|\det\Gamma|$ is sufficiently large.  

For the latter convenience, we first show the following lemma: Let $A$ be a complex matrix.  Then $B\equiv (A+A^\dagger)/2$ is Hermitian and diagonalized by an appropriate unitary matrix $P$ as $P^{\dagger}BP={\rm diag}(\lambda_1^B, \cdots, \lambda_N^B)\equiv\Lambda$, where $^\dagger$ denotes the conjugate transpose, and $\lambda_1^B, \cdots, \lambda_N^B$ the eigenvalues of $B$.  We claim there exists an anti-Hermitian (anti--symmetric) matrix $G$ satisfying 
$\det A=\det(\Lambda+G)$.  
Since $P\Lambda P^\dagger =B$, $G$ is indeed given as 
$G=P^\dagger\frac{A-A^\dagger}{2}P$, 
which is anti-Hermitian when $A$ is a complex matrix.  If $A$ is a real matrix, $G$ and $P$ are real anti--symmetric and orthogonal, respectively.  

\begin{figure}[thbp]
\includegraphics[width=90mm]{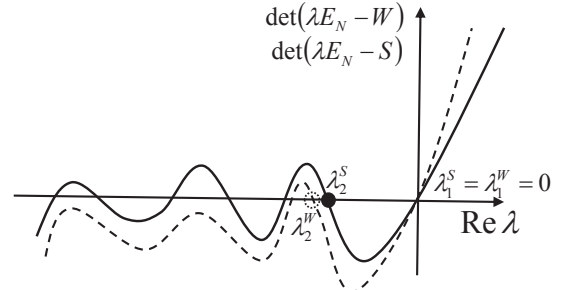}
\caption{Since $\lambda_1^W=\lambda_1^S=0$ and $d\det(\lambda E_N-W)/d\lambda |_{\lambda=0}\ge d\det(\lambda E_N-S)/d\lambda |_{\lambda=0}$, $\det(\lambda E_N-W)$ (the dashed line) is smaller than $\det(\lambda E_N-S)$ (the solid line) for negative $\lambda$ sufficiently close to $\lambda=0$.  Since $|\det(\lambda E_N-W)|=\det({\rm Re }\lambda E_N-S)$ does not hold in the region $\lambda_2^S<{\rm Re }\,\lambda<0$ according to (\ref{eq:I2}), the two lines cannot cross each other in this region.  The filled and dotted circles denote $\lambda_2^S$ and $\lambda_2^W$, respectively.  }\label{fig:1}
\end{figure}
The essence of the proof for proposition (I) is the Ostrowski-Taussky inequality~\cite{Horn}.  Ostrowski-Taussky inequality claims that 
$|\det A|\ge\det\dfrac{A+A^\dagger}{2}$ 
holds if $(A+A^\dagger )/2$ is positive definite, i.e., ${\bm v}^\dagger\frac{A+A^\dagger}{2}{\bm v}>0$ for an arbitrary complex vector ${\bm v}$.  Note that the equality holds if and only if $A$ is Hermitian.  In our case, since the largest eigenvalue of $S$ is zero, the matrix ${\rm Re }\,\lambda E_N - S$ is positive definite if ${\rm Re }\,\lambda>0$.  Then $|\det(\lambda E_N-W)|\ge\det({\rm Re }\,\lambda E_N-S)$ holds for ${\rm Re }\,\lambda>0$ according to the Ostrowski-Taussky inequality.  Since the largest eigenvalues of $W$ and $S$ both are zero, restricting $\lambda$ to be real, this fact reads 
\begin{eqnarray}
\dfrac{d\det(\lambda E_N-W)}{d\lambda}\Big|_{\lambda=0}\ge\dfrac{d\det(\lambda E_N-S)}{d\lambda}\Big|_{\lambda=0}\label{eq:I1}
\end{eqnarray}
(the equality holds if and only if $W$ is symmetric).  Note that $\det(\lambda E_N-W)$ and $\det(\lambda E_N-S)$ both are smooth functions with respect to $\lambda$.  

Next, let us show the proof of proposition (II).  According to the lemma shown above, there exists an anti-Hermitian matrix of order $(N-1)$, $\tilde{\Gamma}$, which satisfies $\det(\lambda E_{N-1}-\tilde{\Lambda} - \tilde{\Gamma})=\det(\lambda E_N-W)/\lambda$, where $\tilde{\Lambda}={\rm diag }(\lambda_2^S, \cdots, \lambda_N^S)$.  Similarly to the discussion in the proof of proposition (I), ${\rm Re }\,\lambda E_{N-1}-\tilde{\Lambda}$ is positive definite if ${\rm Re }\,\lambda>\lambda_2^S$.  Using the Ostrowski-Taussky inequality again, we obtain $|\det(\lambda E_{N-1}-\tilde{\Lambda}-\tilde{\Gamma})|\ge\det({\rm Re }\,\lambda E_{N-1}-\tilde{\Lambda})$, which implies  
\begin{eqnarray}
\left|\dfrac{1}{\lambda}\det(\lambda E_N-W)\right|\ge\dfrac{1}{\lambda}\det({\rm Re }\lambda E_N-S)\label{eq:I2}
\end{eqnarray}
for $\lambda_2^S<{\rm Re }\,\lambda<0$.  The equality holds if and only if $W$ is symmetric.  As a consequence of Eq.~(\ref{eq:I2}), $\det(\lambda E_N-W)=0$ does not have a solution in the region $\lambda_2^S<{\rm Re }\,\lambda<0$.  Therefore ${\rm Re }\,\lambda_2^W-\lambda_2^S\le 0$ always holds (see Fig.~\ref{fig:1}).    

The combination of the above two propositions implies that ${\rm Re }\,\lambda_2^W$ is always smaller than $\lambda_2^S$, which means the system without DBC always converges faster than the system with DBC.  Therefore the advantage of the violation of DBC is rigorously shown.   

\begin{figure}[htbp]
\hspace{-35mm}
\begin{minipage}{0.5\hsize}
\includegraphics[width=80mm]{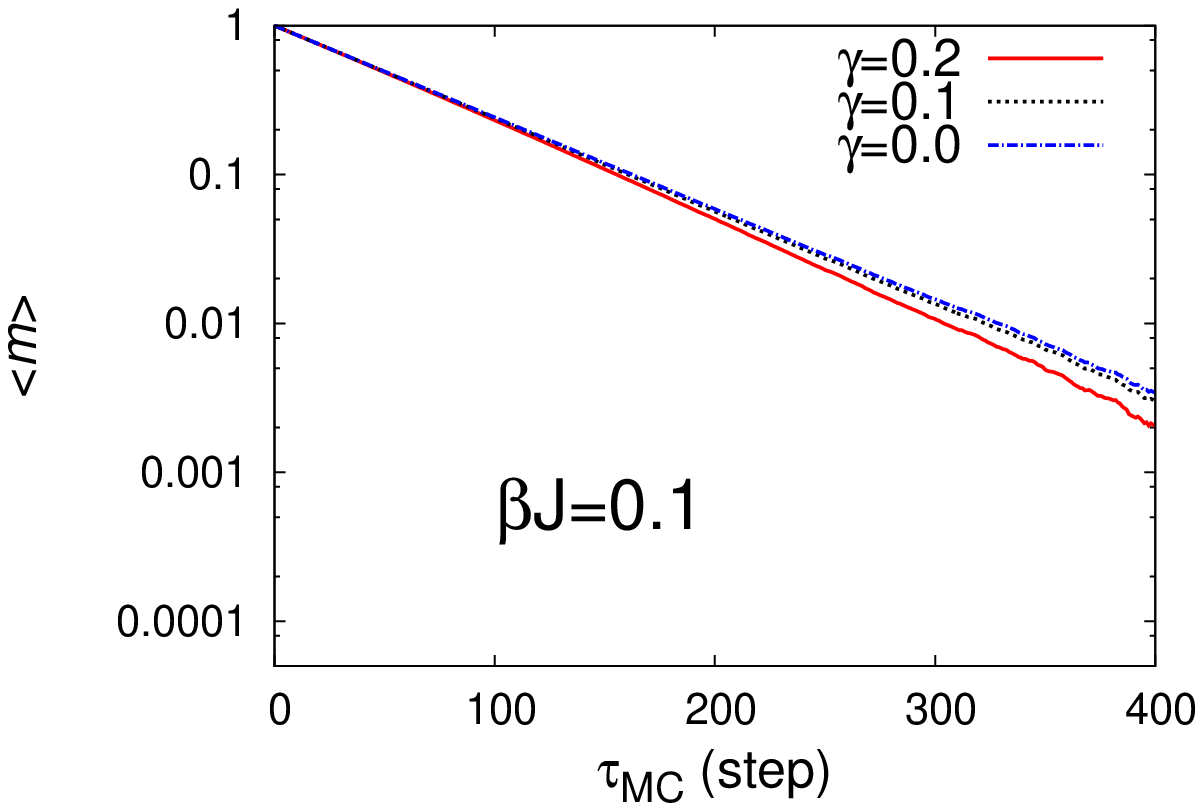}
\end{minipage}\\
\hspace{-35mm}
\begin{minipage}{0.5\hsize}
\includegraphics[width=80mm]{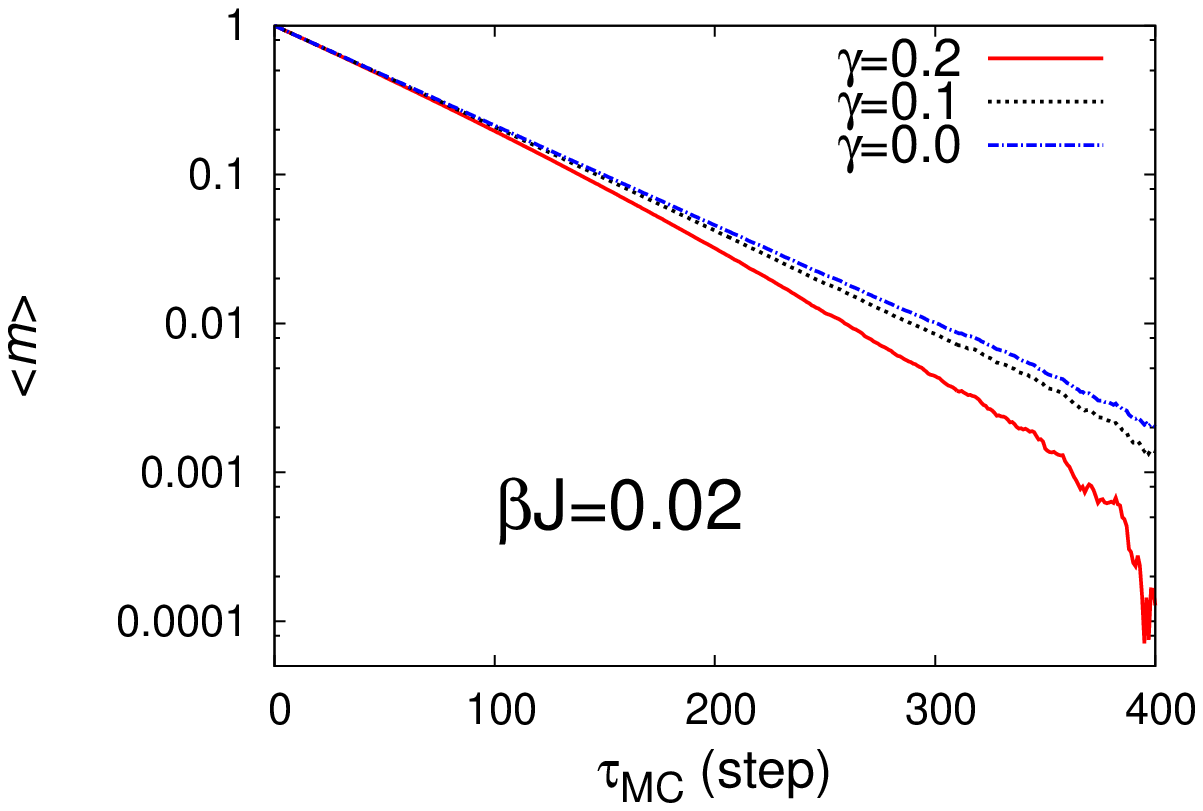}
\end{minipage}
\caption{Relaxations of $\langle m(\tau_{\rm MC})\rangle$ in the 1D Ising model for several values of asymmetry parameter $\gamma$.  The upper and lower figures depict the cases of $\beta J=0.1$ and $0.02$, respectively.}\label{fig:MCMC}
\end{figure}
In order to provide a concrete example of our result, the magnetization of the one-dimensional Ising model with its Hamiltonian $-J\sum_{\langle i,j\rangle}S_i S_j$ has been calculated by asymmetric algorithms, where the summation is taken over all nearest neighbor pairs.  Figure~\ref{fig:MCMC} shows the ensemble average of magnetizations $\langle m(\tau_{\rm MC})\rangle$ as functions of time steps $\tau_{\rm MC}$.  The symmetric part of $W$ is fixed to be one for a Gibbs sampler.  The number of spins is $N=2^7$.  To obtain $\langle m(\tau_{\rm MC})\rangle$, $10^5$ path realizations are averaged.  In order to introduce asymmetry in $W$, double-spin flip for adjacent two spins has been used instead of single-spin flip.  The transition probability for adjacent two spins is given by a $4\times 4$ matrix depending on their nearest neighbor spins.  Correspondingly, the matrix $W$ is truncated to a $4\times 4$ matrix, and its asymmetric part $\Gamma$ has three independent parameters.  For simplicity, we have fixed them as $\Gamma_{++\to +-}=\Gamma_{++\to -+}=\Gamma_{+-\to -+}=\gamma\exp(-4\beta J)$, where $\beta$ is the inverse temperature.  The other components of $\Gamma$ are determined by Eq.~(\ref{eq:BCandProbConserv}).  Note that $|\gamma|\le c\exp(-d\beta J)$ should be satisfied so that all components of the transition matrix represent probability, i.e., $0\le q_{ij}\Delta t+\delta_{ij}\left( 1-\sum_j q_{ji}\Delta t\right)\le 1$, where $c$ and $d$ are determined by system dimension and lattice type, and $\Delta t$ time step.  Because of this restriction to $\gamma$, the difference between symmetric ($\gamma=0$) and asymmetric ($\gamma\neq 0$) algorithms becomes significant for small $\beta J$,namely in high temperature, as shown in Fig.~\ref{fig:MCMC}.  

In this paper, we have proven that the system without DBC always relaxes toward the target distribution faster than the system with DBC.  
The central cue of the proof is given by $W_{ij}$.  This fact implies that the implementation of MCMC should be given by designing $W$.  From the perspective of stochastic quantization, this is equivalent to designing the Hamiltonian, which is non-Hermitian.  Our result shows that the introduction of the anti-symmetric part always decreases the real part of the second-largest eigenvalue.  This fact implies that the speed of nonequilibrium relaxation toward the target (steady state) distribution $\pi_i$ is bounded from below by the properties in the corresponding equilibrium system described by the same equilibrium probability $\pi_i$.  This fact implies that there exists a kind of thermodynamic inequality, which bridges between the relaxation and response in nonequilibrium and equilibrium systems.  Its physical interpretation requires further studies.  In mathematics, it is known that the introduction of a non--conservative driving force, which does not alter the stationary distribution, accelerates the convergence for normal diffusion systems~\cite{Hwang, Hwang2, Lelievre}.  Our result is regarded as the generalization of these results.  

Consequently, there exist two choices to accelerate the convergence of MCMC.  The one is designing the symmetric part $S$.  Since DBC is equivalent to symmetric $W$, this choice is within the framework of DBC.  Examples of such improvement are found in well-known extended ensemble methods and clustering algorithms.  Another choice is given by arranging the anti--symmetric part $\Gamma$.  The hybrid use of these two approaches would improve the convergence of MCMC.  

Since the average reject rate is given by the trace of $W$, the decrease of the reject rate means only the decrease of the sum of all eigenvalues of $W$.  Thus it is not confirmed that the second-largest eigenvalue, which determines the convergence speed, decreases.  Suwa and Todo explained that the acceleration of the convergence in their proposed method is responsible for the reduction of reject rate~\cite{SuwaTodo}.  However, the direct reason for the acceleration is the reduction of the second-largest eigenvalue due to the introduction of the anti--symmetric part $\Gamma$, as shown in this paper.  

Our result is considered to be efficient even for the large $N$ system.  The essence of acceleration is the degeneracy of eigenvalues induced by asymmetry.  As $|\det\Gamma|$ increases, ${\rm Re }\,\lambda_2$ decreases and ${\rm Re }\,\lambda_3$ increases.  In the limiting case, where all eigenvalues except $\lambda_1=0$ degenerate, $\lambda_2={\rm tr }S/(N-1)$.  Thus ${\rm Re }\,\lambda_2\ge {\rm tr }S/(N-1)$ gives the restriction on the acceleration.  Since ${\rm tr }S$ is regarded to be proportional to $N$, ${\rm Re }\,\lambda_2$ is always allowed to shift by order of unity by the introduction of asymmetry.  

Our proof does not show the optimal implementation for fast convergence toward the target distribution.  
The most important problem is which algorithm given by DBC should be compared to that without DBC.  In Ref.~\cite{SuwaTodo}, it is reported that their proposed method shows the convergence more than six times faster than that by the Metropolis algorithm for the Potts model.  However, the algorithm described by the symmetric part $S$ in the Suwa-Todo method has not been specified.  In our framework, it is ensured only that the convergence in the system with anti--symmetric part $\Gamma$ is always faster than that without $\Gamma$.  Thus the comparison between the Suwa-Todo method and the Metropolis algorithm is nonsense from a point of view of violating DBC, which is our standpoint.  It is required to specify the system described by the symmetric part $S$ induced by the DBC-violating algorithms.  The comparison to the corresponding system to the symmetric part is relevant for assessment of the performance of the violation of DBC.  The physical interpretation of the effect of DBC violation requires further understanding.  Such interpretation is addressed in Ref.~\cite{OhzekiIchiki} and other problems will be discussed in our sequel studies.  

The authors thank Yuki Sughiyama and Koji Hukushima for their fruitful discussions.  The authors also acknowledge Yuji Sakai and Hugo Touchette for their useful comments.  This work is partially supported by MEXT in Japan, Grant-in-Aid for Young Scientists (B) Grant No. 24740263.

\end{document}